\documentclass[fleqn,usenatbib]{mnras}

\DeclareRobustCommand{\VAN}[3]{#2}
\let\VANthebibliography\thebibliography
\def\thebibliography{\DeclareRobustCommand{\VAN}[3]{##3}\VANthebibliography}

\usepackage{newtxtext,newtxmath}
\usepackage[T1]{fontenc}
\usepackage{graphicx}	
\usepackage{amsmath}	
\usepackage{comment}
\usepackage{xspace} 
\usepackage{tabu}
\usepackage{multirow}
\usepackage[dvipsnames]{xcolor} 
\graphicspath{{Figures/}}

\defcitealias{DESY1CL}{Y1CL}
\def\beq{\begin{eqnarray}}
\def\eeq{\end{eqnarray}}

\def\pimax{\Pi_\mathrm{max}}

\newcommand{\beqa}{\begin{equation}\begin{aligned}}
\newcommand{\eeqa}{\end{aligned}\end{equation}}
\newcommand{\bit}{\begin{itemize}}
\newcommand{\eit}{\end{itemize}}

\newcommand{\hiMpc}{h^{-1} \rm Mpc}
\newcommand{\hikpc}{h^{-1} \rm kpc}
\newcommand{\hiMsun}{h^{-1} M_\odot}

\newcommand{\OmegaM}{\Omega_\mathrm{M}}

\def\wp{w_\mathrm{p}} 
\newcommand{\rp}{r_\mathrm{p}}
\newcommand{\bh}{b_\mathrm{h}}
\newcommand{\bg}{b_\mathrm{g}}
\newcommand{\bc}{b_\mathrm{c}}
\newcommand{\ximm}{\xi_\mathrm{mm}}
\newcommand{\xigg}{\xi_\mathrm{gg}}
\newcommand{\xicc}{\xi_\mathrm{cc}}

\newcommand{\xihm}{\xi_\mathrm{hm}}

\newcommand{\xicm}{\xi_\mathrm{cm}}
\newcommand{\xihg}{\xi_\mathrm{hg}}
\newcommand{\xicg}{\xi_\mathrm{cg}}

\newcommand{\wpmm}{w_\mathrm{p,mm}}
\newcommand{\wphm}{w_\mathrm{p,hm}}
\newcommand{\wpcm}{w_\mathrm{p,cm}}
\newcommand{\wpcg}{w_\mathrm{p,cg}}

\newcommand{\wpgg}{w_\mathrm{p,gg}}

\newcommand{\DS}{\Delta\Sigma}

\newcommand{\fcen}{f_\mathrm{cen}}
\newcommand{\Mmin}{M_\mathrm{min}}
\newcommand{\siglogM}{\sigma_{\log M}}

\newcommand{\redmapper}{redMaPPer\xspace} 
\newcommand{\redmagic}{redMaGiC\xspace}

\title[Self-calibrating galaxy cluster selection bias]{Self-calibrating optical galaxy cluster selection bias using cluster, galaxy, and shear cross-correlations}

\author[C. Zeng et al.]{Chenxiao Zeng,$^{1,2}$\thanks{E-mail: zeng.544@osu.edu}
Andr\'{e}s N. Salcedo$^{3}$,
Hao-Yi Wu$^{4}$,
and Christopher M. Hirata$^{1,2,5}$
\\
$^{1}$Department of Physics, The Ohio State University, 191 West Woodruff Avenue, Columbus, OH 43210, USA \\
$^{2}$Center for Cosmology and AstroParticle Physics, The Ohio State University, Columbus, OH 43210, USA \\
$^{3}$ Department of Astronomy/Steward Observatory, University of Arizona, 933 North Cherry Avenue, Tucson, AZ 85721-0065, USA \\
$^{4}$Department of Physics, Boise State University, Boise, ID 83725, USA \\
$^{5}$Department of Astronomy, The Ohio State University, 140 West 18th Avenue, Columbus, OH 43210, USA \\
}
\pubyear{2022}
\begin{document}
\label{firstpage}
\pagerange{\pageref{firstpage}--\pageref{lastpage}}
\maketitle

\begin{abstract}

The clustering signals of galaxy clusters are powerful tools for self-calibrating the mass--observable relation and are complementary to cluster abundance and lensing.  In this work, we explore the possibility of combining three correlation functions --- cluster lensing, the cluster--galaxy cross-correlation function, and the galaxy auto-correlation function --- to self-calibrate optical cluster selection bias, the boosted clustering and lensing signals in a richness-selected sample mainly caused by projection effects.  We develop mock catalogues of \redmagic-like galaxies and \redmapper-like clusters by applying Halo Occupation Distribution (HOD) models to N-body simulations and using counts-in-cylinders around massive haloes as a richness proxy.  In addition to the previously known small-scale boost in projected correlation functions, we find that the projection effects also significantly boost 3D correlation functions to scales of 100 $\hiMpc$.  We perform a likelihood analysis assuming survey conditions similar to the Dark Energy Survey (DES) and show that the selection bias can be self-consistently constrained at the 10\% level.  We discuss strategies for applying this approach to real data.  We expect that expanding the analysis to smaller scales and using deeper lensing data would further improve the constraints on cluster selection bias.

\end{abstract}


\section{Introduction}

The abundance of galaxy clusters across cosmic time reflects the growth rate of cosmic structure and is a sensitive probe of cosmic acceleration \citep[see e.g.][]{Frieman08, Allen11, Weinberg13, Huterer15}.  The halo mass function predicts the halo number density as a function of mass and redshift for a given set of cosmological parameters \citep[see e.g.][]{PressSchechter74, Sheth01, Tinker08MF}.  To connect this theoretical prediction with the observed cluster abundance, we need well-calibrated and unbiased mass--observable relations.  The mass--observable relation can be derived from combinations of X-ray luminosity and temperature \citep[e.g.][]{RozoRykoff14, Giles22}, Sunyaev--Zeldovich (SZ) effect \citep[e.g.][]{Saro15, Bleem20}, galaxy velocity dispersion \citep[e.g.][]{Bocquet15, Rozo15RM4}, and weak gravitational lensing \citep[e.g.][]{Melchior17, Simet17, Murata18, Dietrich19, McClintock19, Murata19}.  The accuracy of the mass--observable relation critically impacts the constraining power of the cluster sample \citep[e.g.][]{Wu21}.

Deriving cosmological parameter constraints by combining observed cluster abundances and the mass--observable relation has been the strategy of many previous studies \citep[e.g.][]{Vikhlinin09, Mantz10, Rozo10, Mantz14, Bocquet15, Planck15Cosmo, deHaan16, Bocquet19, DESY1CL, Costanzi21}.  A complementary approach would be to use the clustering of clusters (correlation functions or power spectra) to self-calibrate the mass--observable relation \citep[e.g.][]{LimaHu04, Majumdar04, LimaHu05, Wu08, Salcedo20}. This strategy has been applied to X-ray surveys \citep[e.g.][]{Collins00, Schuecker03, Balaguera11} and optical surveys \citep[e.g.][]{Croft99, Sanchez05, Estrada09, Mana13, Baxter16, Paech17, Chiu20, To21b, Park21}. In particular, recent wide-field optical surveys have enabled precision cosmology analyses using cluster clustering. For example, \cite{To21b} combine the auto- and cross-correlations between clusters, galaxies, and shear from the DES to derive competitive cosmological constraints.

However, recent studies show that optically selected galaxy clusters exhibit selection bias in their lensing and clustering signals \citep{DESY1CL, Sunayama20, Wu22}.  In particular, at a given mass, a richness-selected sample tends to have a higher lensing and clustering signal than expected from their masses.  This selection bias, if not accounted for, will lead to biased cluster mass calibration and cosmological parameters.  This selection bias has been mostly ignored in previous studies but has become one of the dominant systematic uncertainties for current DES data \citep{DESY1CL}.

In this work, we use multiple correlation functions to self-calibrate optical cluster selection bias.  \cite{Salcedo20} have previously shown that combining cluster lensing, cluster--galaxy cross-correlations, and galaxy auto-correlations provides an effective way to break the degeneracy between the scatter in the richness--mass relation and the matter density fluctuation amplitude $\sigma_8$.  The basic idea is that these three observables can be combined to solve for three unknowns: cluster bias $\bc$, galaxy bias $\bg$, and $\sigma_8$.  The resulting $\bc$ directly constrains the scatter. We similarly use these observables to simultaneously solve for $\bg$, $\bc$, and $\sigma_8$; here, $\bc$ includes the effect of selection bias.  We construct mock cluster and galaxy samples by applying the HOD framework \citep[e.g.][]{BerlindWeinberg02, CooraySheth02, Zheng05, Zehavi11} to the {\sc Abacus Cosmos} N-body simulation suite.  We calculate three-dimensional and two-dimensional correlation functions between clusters, galaxies, and matter.  We show that we can correctly recover the cluster bias $b_c$ and $\sigma_8$ by fitting these correlation functions simultaneously.  We focus on scales greater than $10~\hiMpc$ and defer small-scale calibration to future work.  This work paves the way for an analysis using wide-field survey data like DES and Vera C.~Rubin Observatory Legacy Survey of Space and Time (LSST).

This paper is organised as follows. Section ~\ref{sec:dataset} describes the simulated mock catalogues, and Section~\ref{sec:clutering_stats} describes our measurements of correlation functions.  In Section \ref{sec:mcmc}, we present the likelihood analysis for self-calibrating cluster selection bias and constraining cosmological parameters. We discuss our results in Section~\ref{sec:discussion} and summarise in Section~\ref{sec:summary}. In this work, we use the fiducial flat {\em Planck} $\Lambda$CDM cosmology \citep{Planck15Cosmo} adopted by the {\sc Abacus Cosmos} simulation suite: $\OmegaM$ = 0.314, $h$ = 0.673, $\sigma_8$ = 0.83, $n_s$ = 0.9652, $\Omega_{\rm B}$ = 0.049.  All distances are in comoving $\hiMpc$.
We use the spherical overdensity mass definition $M_\mathrm{200m}$, defined such that the mean density enclosed is 200 times the mean density of the Universe.

\section{Mock Galaxy and Cluster Catalogues}
\label{sec:dataset}

We generate mock galaxy catalogues by applying HOD models to the {\sc Abacus Cosmos} N-body simulations.  We apply two sets of HOD parameters. The first one simulates \redmagic galaxies, which have   precise photometric redshifts and are optimised for calculating cross- and auto-correlation functions (Table~\ref{fig:HOD_redmagic}).  The second one simulates the members of \redmapper clusters  (Table~\ref{fig:HOD_redmapper}).  Below we describe our approach in detail.

\subsection{Abacus Cosmos N-body simulations}

We build our mock galaxy and cluster catalogues using the public {\sc Abacus Cosmos} N-body simulation suite\footnote{\href{https://lgarrison.github.io/AbacusCosmos/}{https://lgarrison.github.io/AbacusCosmos/}} \citep{Garrison18}, which is based on the {\sc Abacus} N-body code \citep{Metchnik09, Garrison18}.   We use 20 periodic boxes of the fiducial {\em Planck} cosmology with varied phases in the initial conditions, with a box size 1100 $\hiMpc$ (internally called $\tt AbacusCosmos\_1100box\_planck$).  We focus on the $z=0.3$ outputs in this work.  Each simulation box contains 1440$^3$ dark matter particles, corresponding to a mass resolution of $4\times 10^{10}~\hiMsun$, and has a spline softening of 63 $\hikpc$.  Dark matter halo catalogues are created by applying the {\sc Rockstar} halo finder \citep{Behroozi13rs} to particle snapshots.  For assigning galaxies to haloes, we use host haloes defined by {\sc Rockstar} and the mass definition $M_{\rm 200 m}$.  For the lensing calculations, we use a 0.1\% subsample of the dark matter particles, and we have tested that this downsampling can accurately recover the lensing signal well below 0.1 $\hiMpc$.

\subsection{Mock \redmagic sample}

\begin{table}
   \centering
   \caption{Fiducial values and descriptions for HOD parameters for the mock \redmagic galaxy catalogues.}
    \begin{tabular}{ccl}      
    \hline
    Parameter & Fiducial & Description \\
    \hline
    $\sigma_{\log M}$ &  $0.60$ & width of central transition\\
    $\log M_\mathrm{min}$ & $12.7$ & minimum halo mass to host a central\\
    $\log M_{0}$& $11.0$ & satellite cut-off mass\\
    $\log M_{1}$ & $13.8$ & minimum halo mass to host a satellite\\
    $\alpha$ & $1.50$ & slope of satellite occupation power law\\
    $f_\mathrm{cen}$ & $0.60$ & central completeness fraction\\
    \hline
   \end{tabular}
\label{fig:HOD_redmagic}
\end{table}
\begin{table}
   \centering
   \caption{Similar to Table~\ref{fig:HOD_redmagic}, but for galaxies that match the colour--magnitude selection of the member galaxies of \redmapper clusters.  This HOD is different from that for \redmagic galaxies because they correspond to different selection criteria.}
    \begin{tabular}{ccl}      
    \hline
    Parameter & Fiducial & Description \\
    \hline
    $\log M_\mathrm{min}$ & $12$ & mass threshold of haloes\\
    $\log M_{0}$& $11.7$ & same as in Table~\ref{fig:HOD_redmagic}\\
    $\log M_{1}$ & $12.9$ & same as in Table~\ref{fig:HOD_redmagic}\\
    $\alpha$ & $1$ & same as in Table~\ref{fig:HOD_redmagic}\\
    \hline
   \end{tabular}
\label{fig:HOD_redmapper}
\end{table}

The \redmagic galaxy samples \citep{Rozo16} are designed to minimise photometric redshift uncertainties and have been used in various galaxy clustering and lensing studies \citep[e.g.][]{DESY1KP}. We populate simulated {\sc{Abacus Cosmos}} haloes with mock DES \redmagic galaxies using an HOD model. As in \citet{Salcedo22}, we extend this framework to include central incompleteness, which is known to affect \redmagic galaxies due to the strict colour selection criteria.  We parameterise the mean central and satellite occupations of our haloes as

\begin{align}
\langle N_\mathrm{cen} | M_h \rangle &= \frac{\fcen}{2} \left[ 1 + \mathrm{erf} \left( \frac{\log M_h - \log \Mmin}{\siglogM} \right) \right] \ ,
\label{eq:Nsat}
\\
\langle N_\mathrm{sat} | M_h \rangle &= \frac{\langle N_\mathrm{cen} | M_h \rangle}{\fcen} \left( \frac{M_h - M_0}{M_1} \right)^{\alpha} \ ,
\label{eq:Ncen}
\end{align}

where $\fcen$ allows for central incompleteness, i.e. the fact that not all high-mass haloes have a central satisfying the \redmagic selection criteria. Table~\ref{fig:HOD_redmagic} lists the fiducial values we assume for each of our parameters.

The number of mock central and satellite galaxies placed into each halo is drawn randomly from a binomial and Poisson distribution respectively with mean occupations as given above. Centrals are placed at the centre of their host halo while satellites are distributed according to a Navarro--Frenk--White profile \citep[NFW;][]{Navarro97} parameterised by halo concentration $c_{\rm 200m}$ assigned using the fits of \citet{Correa15}. The extent to which satellite galaxies trace their host halo's dark matter profile is an open question, but because our analysis only relies on large scales ($r > 10 \; \hiMpc$) our results are unaffected by the assumption that  galaxy and halo concentrations are the same.

\subsection{Mock \redmapper sample}

\begin{figure}
\centering
\includegraphics[width=1.\columnwidth]{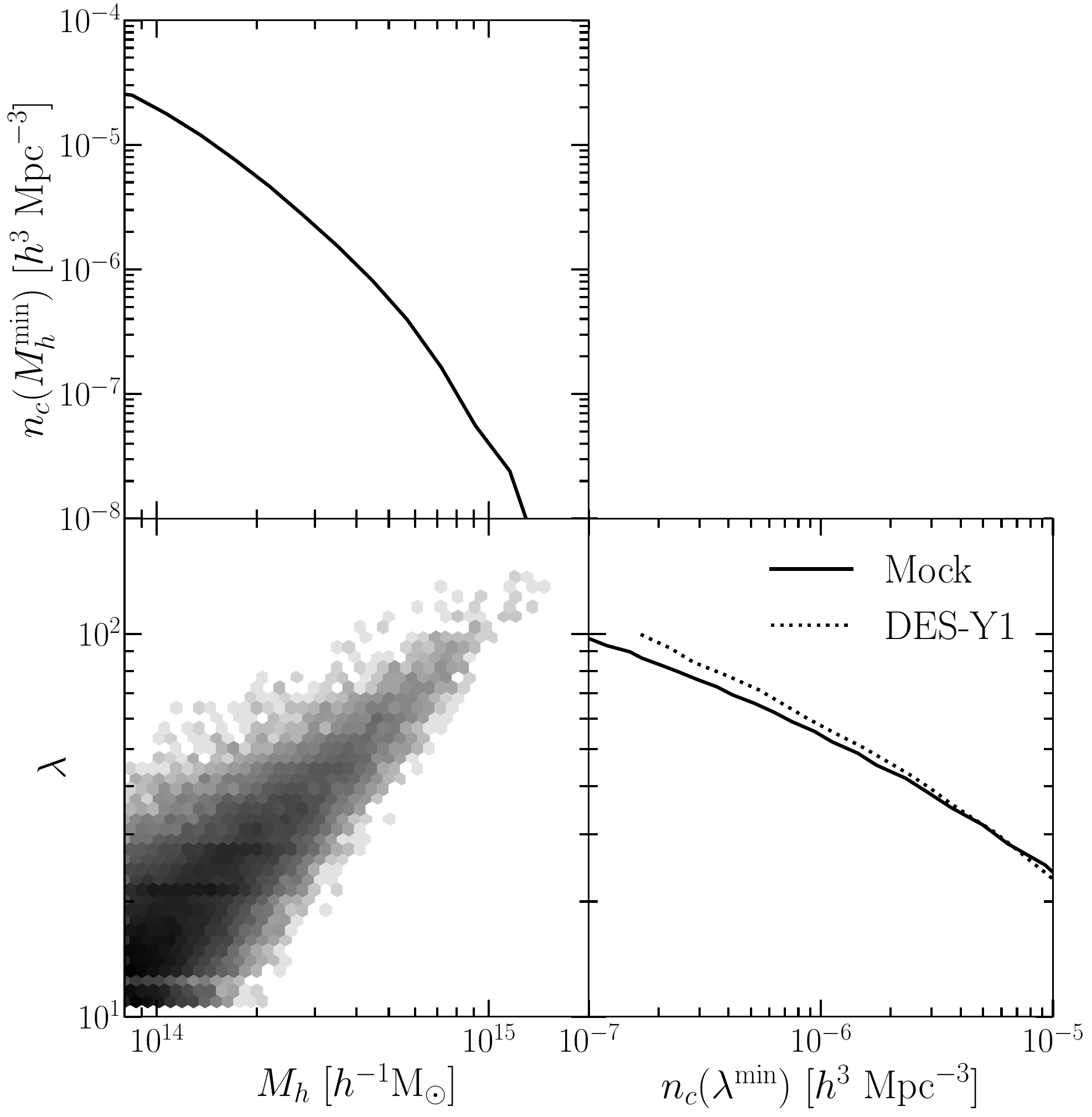}
\caption[]{Richness--mass relation for one of our mock cluster catalogues based on HOD and counts-in-cylinder cluster finding. The top and right-hand panels show the cumulative number density above a given mass and richness, respectively. The cluster abundances from our mock catalogues are broadly consistent with that from DES Y1 cluster sample \citepalias{DESY1CL}, shown as the dotted curve in the right-hand panel.} 

\label{fig:mass-richness-unit}
\end{figure}

The \redmapper cluster finding algorithm \citep{Rykoff14, Rykoff16} identifies clusters from multi-band photometric galaxy catalogues by searching for overdense regions of red galaxies.  The algorithm first trains the red sequence --- the tight relation between colour and magnitude for galaxies in clusters --- as a function of redshift.  The algorithm then uses this red-sequence model to calculate the probability that a galaxy is a member of a potential cluster centre.  The sum of the membership probabilities is the richness of a cluster.

At the beginning of the iteration, all red galaxies are considered potential cluster centres.  After each iteration, potential cluster centres are ranked by their richness values.  A galaxy near multiple cluster centres has a higher priority to be counted toward a higher-ranked centre (a process called `percolation').  The algorithm iterates this process until the resulting cluster catalogue converges.

In principle, one could apply the \redmapper algorithm to a mock galaxy catalogue; for example, it has been applied to the Buzzard simulations \citep{DeRose19} and the CosmoDC2 simulation \citep{Kovacs22}.  However, such a calculation is expensive and requires us to simulate accurate galaxy colours. Therefore, in this work, we simulate the \redmapper catalogue using a simplified counts-in-cylinders approach introduced by \cite{Costanzi19projection} and \cite{Sunayama20}.  

In the first step of our mock \redmapper algorithm, we simulate the `parent population' of \redmapper member galaxies --- galaxies with colours consistent with the \redmapper clusters' red sequence and have the potential to be identified as cluster members.  This step is similar to the initial colour and magnitude selection of the \redmapper algorithm.  These galaxies can contribute to cluster members if they are near the line of sight of a massive halo.

We adopt the HOD parameterisation in \cite{Sunayama20} for this parent population. We assign a central to each halo above $M_\mathrm{min} = 10^{12}\hiMsun$.  For satellite galaxies, we assume
\begin{equation}
\langle N_\mathrm{sat} | M_h \rangle = \langle N_\mathrm{cen} | M_h \rangle \left( \frac{M_h - M_0}{M_1} \right)^{\alpha}  \ ,
\end{equation}
and we list the fiducial values in Table~\ref{fig:HOD_redmapper}.
We emphasise that the HOD model for the \redmapper parent population is different from that of the \redmagic galaxies introduced earlier.  Although \redmagic galaxies and \redmapper member galaxies are both red, they have different colour and magnitude selection criteria and thus different HODs.

In the second step, we mimic the \redmapper cluster finding procedure by counting galaxies within a cylinder along the line of sight.  We use a cylinder depth of $\pm 30 \hiMpc$ (comoving distance along the line of sight). In \cite{Wu22}, we have shown that this projection depth well describes the projection effects and selection bias of \redmapper in the Buzzard simulations.

We assume that each galaxy can only be a member of a single cluster; that is, when a galaxy falls in the cylinders of multiple haloes, it is counted as a member of the most massive one.   This simulates the percolation process of \redmapper.  The resulting number of galaxies inside a cylinder is our mock richness $\lambda$.  The aperture of the cylinder is calculated iteratively based on $\lambda$:

\begin{equation}
    R_\lambda = \left( \frac{\lambda}{100} \right)^{0.2} \quad \mbox{physical $\hiMpc$} \ .
\end{equation}

Fig.~\ref{fig:mass-richness-unit} presents the richness--mass relation of one of our mock \redmapper catalogues (phase 0).  The hexagonal binning presents the number density of haloes in each richness--mass cell. We show the cumulative number density as a function of the mass threshold (top panel) and the richness threshold (right-hand panel). In the right-hand panel, we add the cumulative cluster number density vs.~richness from the DES Y1 \redmapper catalogue\footnote{The DES Y1 \redmapper catalogue is publicly available at \\
\href{https://des.ncsa.illinois.edu}{https://des.ncsa.illinois.edu}.} \citep[][Y1CL thereafter]{DESY1CL}.  The catalogue covers 1437 deg$^2$, and we focus on clusters in the redshift range $0.2 < z < 0.35$.  We assume $\OmegaM = 0.3$ when converting cluster counts to comoving density in the unit of comoving $h^{3} \rm Mpc^{-3}$.  As can be seen, our mock cluster catalogue has a cluster abundance similar to that of the DES Y1 \redmapper catalogue.

For the correlation function calculations, we define a mass-selected halo sample and a richness-selected cluster sample.  For the former, we focus on haloes with $M_{\rm 200m} \ge 2 \times 10^{14} ~ \hiMsun$. This threshold corresponds to approximately 7500 haloes per simulation box of 1100 $\hiMpc$ and a number density $\approx 5.8 \times 10^{-6} ~ h^3 {\rm Mpc}^{-3}$, which corresponds to a richness $\lambda \approx 30$ in \citetalias{DESY1CL}. We then define a richness-selected sample by abundance matching; that is, we sort clusters by their richness values and select the top $N$ clusters that match this number density.

\section{Cluster correlation function observables}
\label{sec:clutering_stats}

With the mock cluster and galaxy catalogues, we are ready to calculate various two-point correlation functions.  Below we briefly introduce the basics of two-point correlation functions and describe our measurements. 

\subsection{Basics for correlation functions}

The two-point cross-correlation function between two sets of points A and B, $\xi_{\rm AB}(r)$, is defined in terms of the joint probability $\delta P$ of finding objects in two volume elements $(\delta V_{\rm A}, \delta V_{\rm B})$ separated by some distance $r$,
\beq
\delta P = n_{\rm A} ~ n_{\rm B} ~ \delta V_{\rm A} ~ \delta V_{\rm B} ~ \left[1 + \xi_{\rm AB}(r) \right],
\eeq
where $n_{\rm A}$ and $n_{\rm B}$ are the respective number densities of sets A and B \citep{Peebles_1980}. The correlation function represents the excess in spatial clustering of sets A and B relative to two uncorrelated sets of points.  Since we use periodic simulations boxes, the correlation functions can be accurately obtained by the natural estimator:

\beq
\xi_{\rm AB}(r) = \frac{{\rm AB}(r)}{{\rm RR}(r)} - 1\ ,
\eeq

where ${\rm AB}(r)$ is the number of A--B pairs with separation $r$, and ${\rm RR}(r)$ is the expected number of pairs in random samples with the same respective number densities and volume geometry.  We calculate the $RR(r)$ analytically using the number densities of $A$ and $B$.

We first calculate the three-dimensional galaxy auto-correlation function $\xigg$, cluster--galaxy cross-correlation function $\xicg$, and cluster--matter cross-correlation function $\xicm$. We use the {\sc Corrfunc} software package \citep{Corrfunc} with 30 logarithmically spaced bins $r$ between 0.1 and 100 $\hiMpc$.  Fig.~\ref{fig:xi} shows our measurements of various $\xi(r)$ functions averaged over 20 mock catalogues.

The projected correlation function is related to the 3D correlation function via

\beq
w_{\rm p,AB}(\rp) = 2 \int_{0}^{\pimax} \xi_{\rm AB}(\rp , \pi) d \pi \ ,
\label{eqn:wp_xi}
\eeq

where $\pi$ is the line-of-sight distance, $\pimax$ is the integration limit along the line of sight, and $\rp$ is the projected distance perpendicular to the line of sight.

Parallel to $\xi$,
we compute the following projected correlation functions: galaxy--galaxy $\wpgg$, cluster--galaxy $\wpcg$, and cluster--matter $\wpcm$.  We note that $\wpcm$ is related to cluster weak lensing signal $\DS$ via a linear transformation \citep[see e.g.][]{Park21local}.

We use {\sc{Corrfunc}} to compute $\xigg(\rp, \pi)$, $\xicg(\rp, \pi)$, and $\xicm(\rp, \pi)$ in 30 logarithmically spaced bins between $0.1 < \rp < 100 \, \hiMpc$, and in linearly spaced $\pi$ bins with $\Delta\pi = 1 \, \hiMpc$ out to $\pimax = 100 \, \hiMpc$. We then sum over the $\pi$ bins to obtain $\wp(\rp)$.  Fig.~\ref{fig:wp} shows our measurements of various $\wp(\rp)$ functions averaged over 20 mock catalogues.

In this work, we use the true 3D positions of galaxies and clusters and do not simulate the photometric redshift uncertainties.  For DES, the redshift uncertainties of \redmapper clusters are $\sigma_z / (1+z) \approx 0.01$ \citep{Rykoff16}, and those of \redmagic galaxies are $\sigma_z / (1+z) \approx 0.017$ \citep{Rozo16}.  The former is likely to be negligible, while the latter can be approximated by a Gaussian distribution.  In general, the impact of photometric redshift errors is to suppress two-point correlation functions consistently across all scales. For a detailed treatment for photometric redshift uncertainties, we refer readers to \cite{ZWang_et_al_2019}.

\subsection{Self-calibrated selection bias}
\label{subsec:self-calibration}

\begin{figure*}
\centering
\includegraphics[width=1.0\columnwidth]{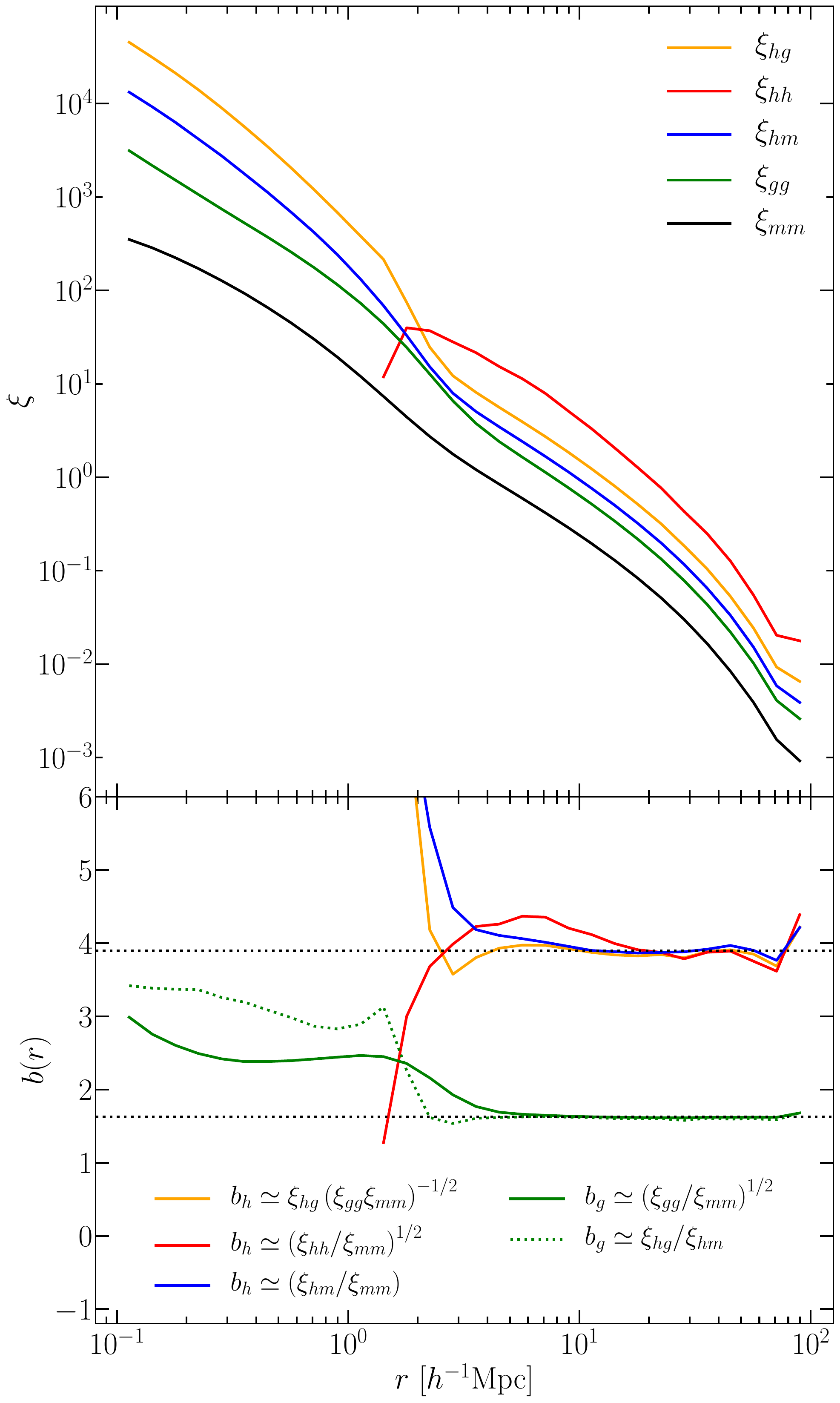}
\includegraphics[width=1.0\columnwidth]{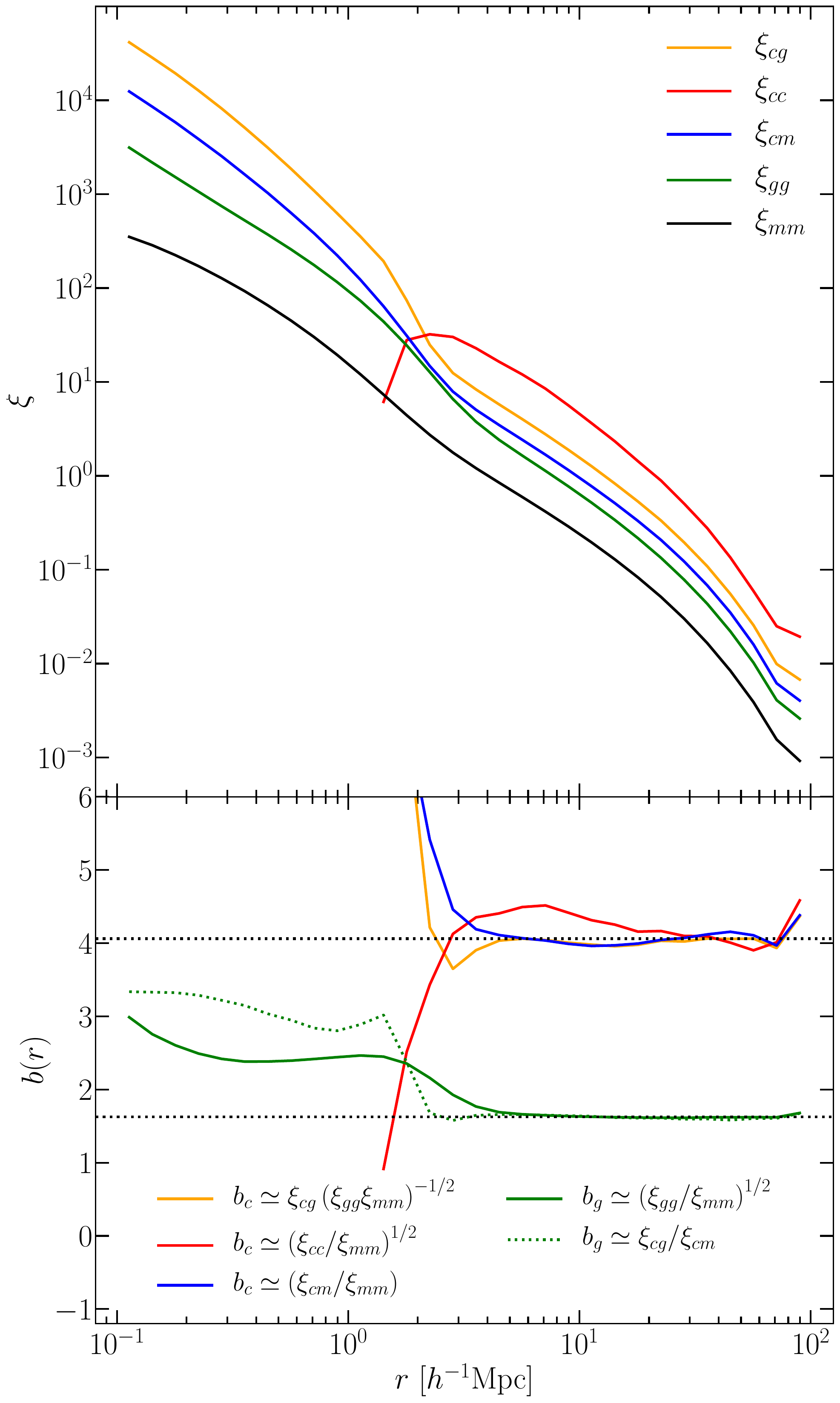}
\caption[]{Top:
three-dimensional two-point correlation functions $\xi(r)$ for different pairs of objects, calculated by averaging over 20 mock catalogues.
Bottom: halo/cluster bias and galaxy bias solved by combining various correlation functions.
Left: haloes above $2\times 10^{14}~\hiMsun$.  The average halo bias $\bh$ and galaxy bias $\bg$ at scales larger than $10~\hiMpc$ are $3.89$ and $1.62$, shown as the horizontal dotted lines.  Right: richness-selected clusters with the same number density as the left panel. The average cluster bias $\bc$ and $\bg$ at scales larger than $10~\hiMpc$ are $4.06$ and $1.63$. 
}
\label{fig:xi}
\end{figure*}

We start by examining the self-consistency between the 3D correlation functions $\xi(r)$ between clusters, galaxies, and matter.  We then use the 2D projected correlation functions $\wp(\rp)$ to self-calibrate selection bias at scales greater than 10 $\hiMpc$.

The left-hand panel of Fig.~\ref{fig:xi} shows the 3D correlation functions for haloes above the $2\times10^{14}~\hiMsun$ threshold, and the right-hand panel shows the analogous calculations for richness-selected clusters with the same number density.  We measure the correlation functions between 0.1 and 100 $\hiMpc$ and average over the 20 phases in {\sc{Abacus Cosmos}}.  The halo auto-correlation curve starts from 2 $\hiMpc$ because of the halo exclusion effects on small scales.

In the lower panel, we show the halo bias $\bh$ and galaxy bias $\bg$ using various combinations of correlation functions. The green solid and dotted curves correspond to galaxy bias $\bg$ computed as,
\beqa
    \bg &= \left( \frac{\xigg}{\ximm} \right)^{1/2}, \\
    \bg &= \frac{\xihg}{\xihm},
\eeqa
while red, blue and orange curves correspond to halo bias $\bh$ computed as,
\beqa
    \bh &= \left( \frac{\xigg}{\ximm} \right)^{1/2},\\
    \bh &= \frac{\xihm}{\ximm},\\
    \bh &=  \frac{\xihg}{\left(\xigg\ximm \right)^{1/2}} \ .
\eeqa
These combinations of correlation functions are largely consistent with each other at large scales, demonstrating the validity of the linear bias model.  The bias calculated using the halo autocorrelation function (red) mildly disagrees with the bias calculated using cross-correlations below $\approx \, 20 \, \hiMpc$ due to halo exclusion.

Averaging over the solid green curves above 10 $\hiMpc$, we obtain $\bg\approx 1.62$. The bias is approximately scale-independent above 2 $\hiMpc$. Similarly, averaging over the blue and orange above 10 $\hiMpc$ gives $\bh \approx 4$. In the right panels, we apply the same procedure to cluster observables. We add horizontal dotted lines to indicate the large-scale bias values.

\begin{figure*}
\centering
\includegraphics[width=1.0\columnwidth]{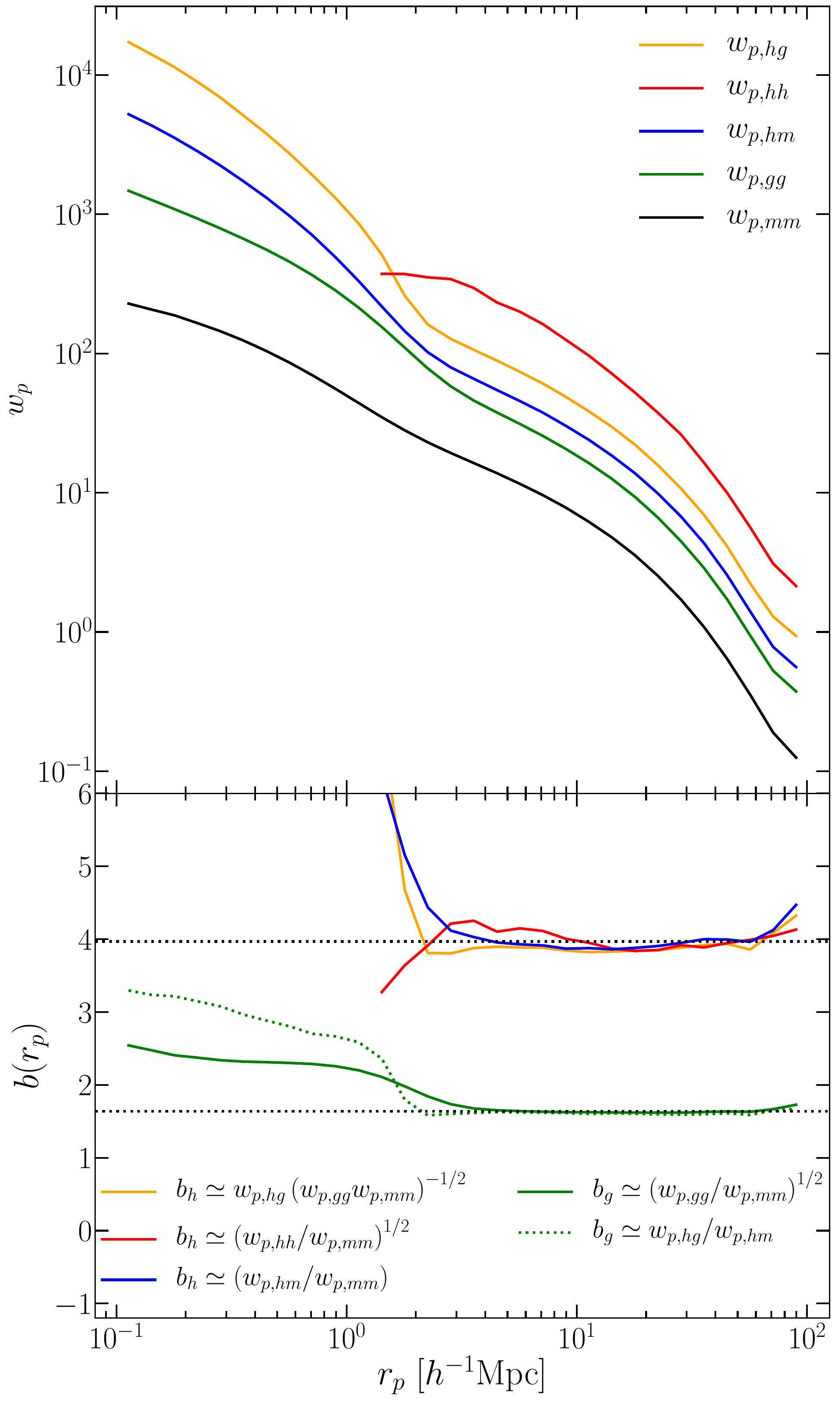}
\includegraphics[width=1.0\columnwidth]{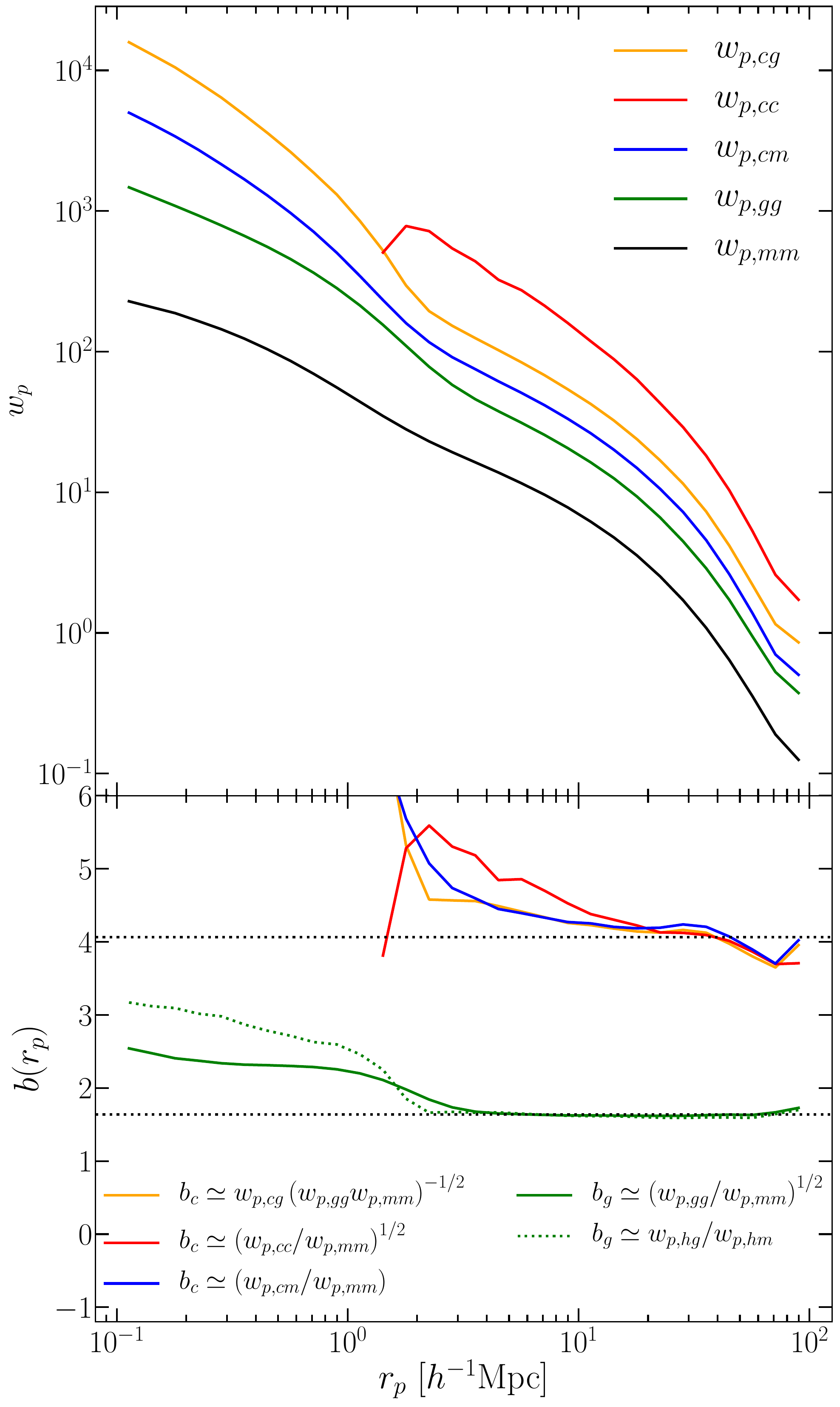}
\caption[]{
Analogous to Fig.~\ref{fig:xi} but for projected correlation functions $\wp(\rp)$. 
Left: haloes.  The average halo bias $\bh$ and galaxy bias $\bg$ at scales larger than $10~\hiMpc$ are $3.97$ and $1.64$, shown as the horizontal dotted lines.
Right: clusters. The average cluster bias $\bc$ and $\bg$ at scales larger than $10~\hiMpc$ are $4.07$ and $1.73$.}
\label{fig:wp}
\end{figure*}

We apply the same procedure (mass cut, abundance matching, and {\sc Corrfunc} pair counts) to calculate the projected correlation functions $\wp(\rp)$, which are presented in Fig.~\ref{fig:wp}. 

From Figs.~\ref{fig:xi} and \ref{fig:wp}, we obtain 4 halo bias and cluster bias values (also summarised in Table~\ref{tab:bias}):
\begin{itemize}
    \item $b_{\rm h,3D}$: 3.89 (from $\xi$) 
    \item $b_{\rm c,3D}$: 4.06 (from $\xi$)
    \item $b_{\rm h,2D}$: 3.97 (from $\wp$)
    \item $b_{\rm c,2D}$: 4.07 (from $\wp$)
\end{itemize}
The halo bias values inferred from $\xi$ and $\wp$ ($b_{\rm h,3D}$ and $b_{\rm h,2D}$) are consistent with each other, indicating that our calculations are self-consistent and that the linear bias model can be applied to our projected correlation functions. We also include the mean mass of our halo and cluster samples averaged over all 20 phases in Table~\ref{tab:bias}. The cluster sample has a lower mean mass due to the scatter in the richness--mass relation and the steepness of the mass function.

\begin{table}
	\centering
	\caption{Summary of the halo and cluster bias values derived from $\xi$ and $\wp$, and the mean halo mass, averaging over 20 realisations of the mock catalogues.  The cluster sample has a lower mean mass but a higher bias, indicating the impact of selection bias.}
	\label{tab:bias}
	\begin{tabular}{ccc}
		\hline
		property & haloes & clusters \\ \hline
		3D bias from $\xi$      & 3.89 & 4.06  \\
		2D bias from $\wp$      & 3.97 & 4.07 \\
		mean mass $[10^{14}~\hiMsun]$ & 3.28 & 2.99 \\\hline
	\end{tabular}
\end{table}

Focusing first on the halo bias and cluster bias from $\xi$ ($b_{\rm h,3D}$ and $b_{\rm c,3D}$), we observe that $b_{\rm h,3D} < b_{\rm c,3D} $.  Given that the latter has a lower mean mass, we would expect a lower bias. However, our results show that the latter has a higher bias despite the lower mean mass.  This indicates that the cluster selection bias is already present in the 3D, non-projected correlation functions.  Our mock cluster catalogues are constructed by counts-in-cylinders along the line of sight, and therefore we expect a line-of-sight boost in the correlation function.  The spherically-averaged 3D correlation function does not eliminate this boost and still exhibits selection bias.

Turning our attention to the cluster bias values inferred from $\xi$ and $\wp$ ($b_{\rm c,3D}$ and $b_{\rm c,2D}$) we  observe that they are roughly equal.  Given that $\wp$ is a projected quantity, one would expect that the line-of-sight boost impacts $\wp$ more strongly than $\xi$.  However, we find that the projection effects boost both similarly.  This suggests that the boosted signal from projection effects is due to a correlation between 3D density and richness that propagates into measurements of 2D clustering.

In our previous work \citep{Wu22}, we have used the Buzzard simulations to calibrate the impact of optical selection bias on $\wpcm$ and $\xicm$. There, we have shown that both $\wpcm$ (equivalent to the surface mass density $\Sigma$) and $\xicm$ (equivalent to the 3D mass density $\rho$) exhibit strong selection bias at $\sim 1 ~\hiMpc$ but have vanishing selection bias at large scales.  In contrast, this work expands to a much larger scale and includes a much larger cluster sample, and we find a non-vanishing boost at large scales and in 3D correlation functions.  We will discuss this point further in Section~\ref{sec:discussion}.

\subsection{Consistency between correlation functions}

\begin{figure}
\centering
\includegraphics[width=1.0\columnwidth]{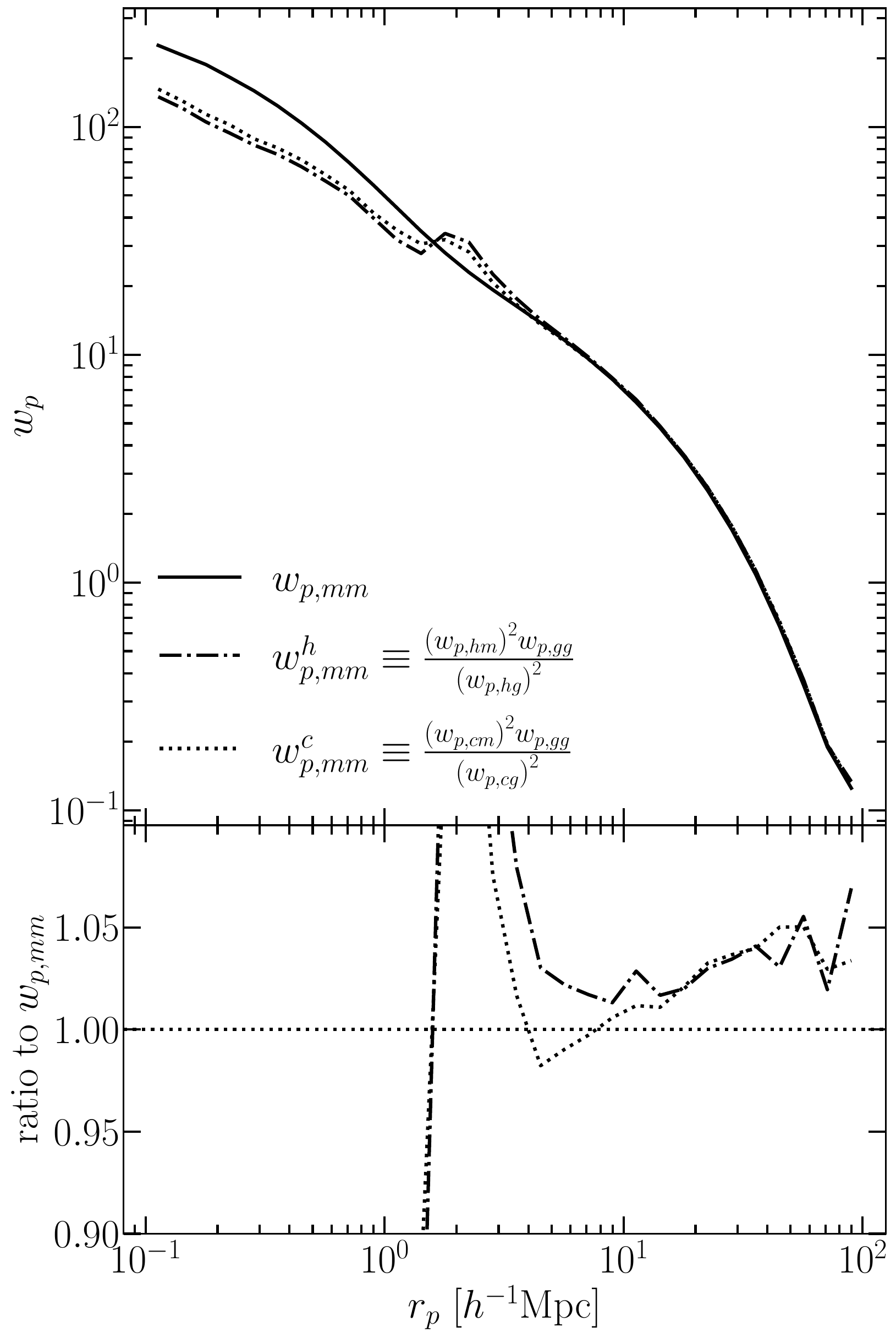}
\caption[]{Projected matter auto-correlation function $\wpmm$ derived from the combination of halo/cluster and galaxy correlation functions.  The dash-dotted curve comes from halo correlation functions, and the dotted curve comes from cluster correlation functions. The solid curve is calculated with the dark matter particles from the simulations. 
The bottom panel shows the ratio of the $\wpmm$ derived from observables with respect to the solid curve.  The excess of $\wpmm $ derived from observables is much smaller than the uncertainty level of DES.
}
\label{fig:wp_compare}
\end{figure}

\begin{figure}
\centering
\includegraphics[width=1.\columnwidth]{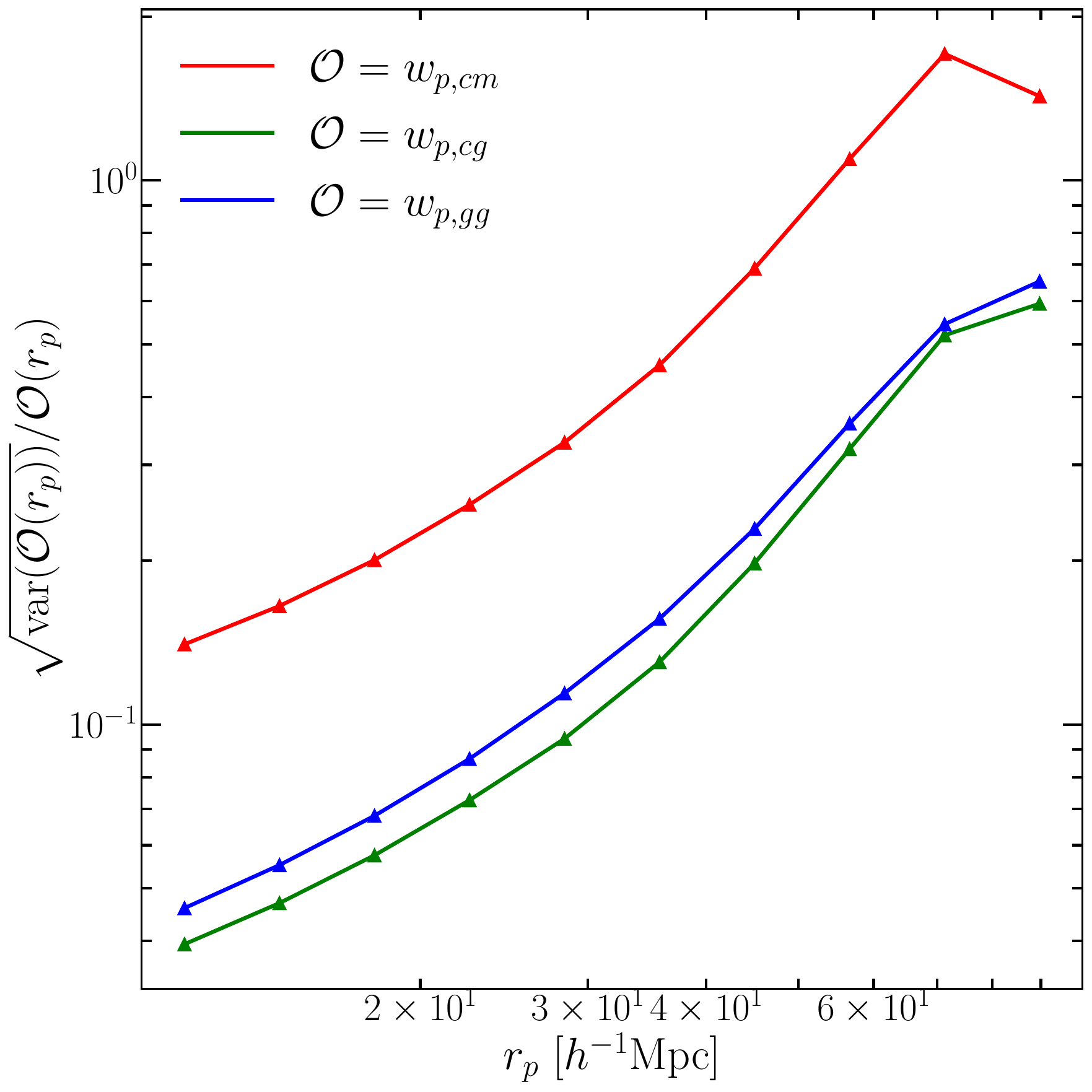}
\caption[]{
Fractional uncertainties for the three correlation functions, assuming a DES-like survey condition: 5000 deg$^2$, for clusters at $0.2 < z < 0.35$.  The lensing noise is contributed by shape noise and large-scale structure noise and is the dominating uncertainty in our likelihood analysis.}
\label{fig:frac_err}
\end{figure}

Before fitting for the parameters, we verify that the correlation functions derived from mock data are consistent with the theoretical expectation. Fig.~\ref{fig:wp_compare} compares the $\wpmm$ calculated directly from simulation particles (solid) with $\wpmm$ derived from galaxy and cluster catalogues assuming linear bias  (dotted and dash-dotted).  We can see that the $\wpmm$ functions derived from the observables have an excess of $\lesssim 5\%$ at scales larger than 10 $\hiMpc$.  This excess is much smaller than the uncertainty levels of DES (see Fig.~\ref{fig:frac_err}).  Therefore, we expect that combining these three correlation functions would allow us to constrain cosmological parameters together with the bias parameters self-consistently.

\section{Likelihood Analysis}
\label{sec:mcmc}

\begin{figure}
\centering
\includegraphics[width=1.\columnwidth]{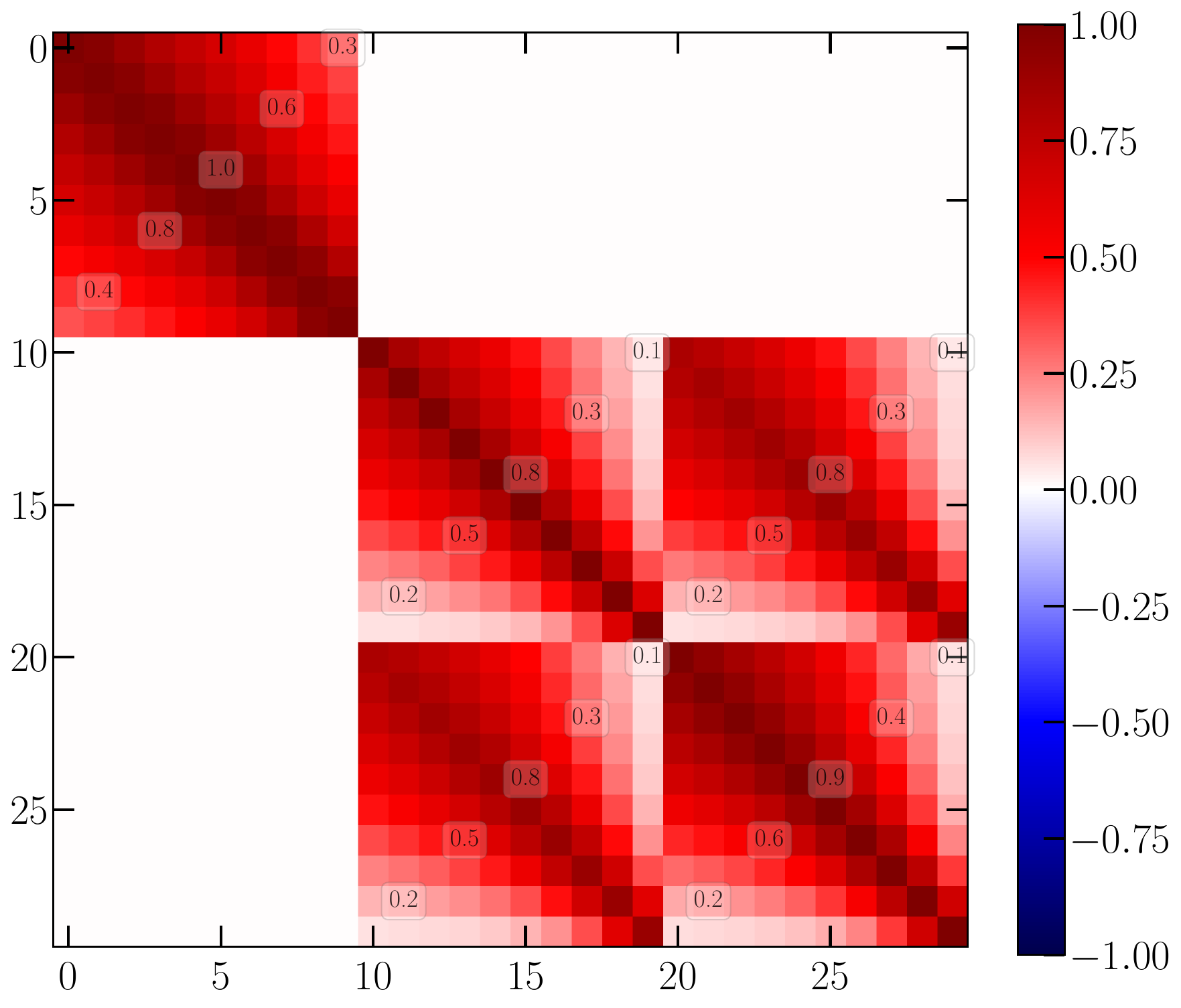}
\caption[]{
Correlation matrix for the three correlation functions, assuming a DES-like survey condition. Bin 0 to 9 (inclusive) refers to $\wpcm$, 10 to 19 refers to $\wpcg$, 20 to 29 refers to $\wpgg$.  We assume $\wpcm$ is uncorrelated with $\wpcg$ and $\wpgg$ (corresponding to the white off-diagonal block) since the $\wpcm$ is dominated by the lensing shape noise and the foreground and background structure uncorrelated with our redshift bin.}
\label{fig:corr}
\end{figure}

Having verified the accuracy of the linear bias model for projected correlation functions at scales larger than $10~\hiMpc$, we perform a likelihood analysis using three $\wp$ functions to constrain the cosmological parameters ($\OmegaM$ and $\sigma_8$) and bias parameters ($\bc$ and $\bg$) simultaneously.

\subsection{Data vector and covariance matrix}

\begin{figure*}
\centering
\includegraphics[width=2\columnwidth]{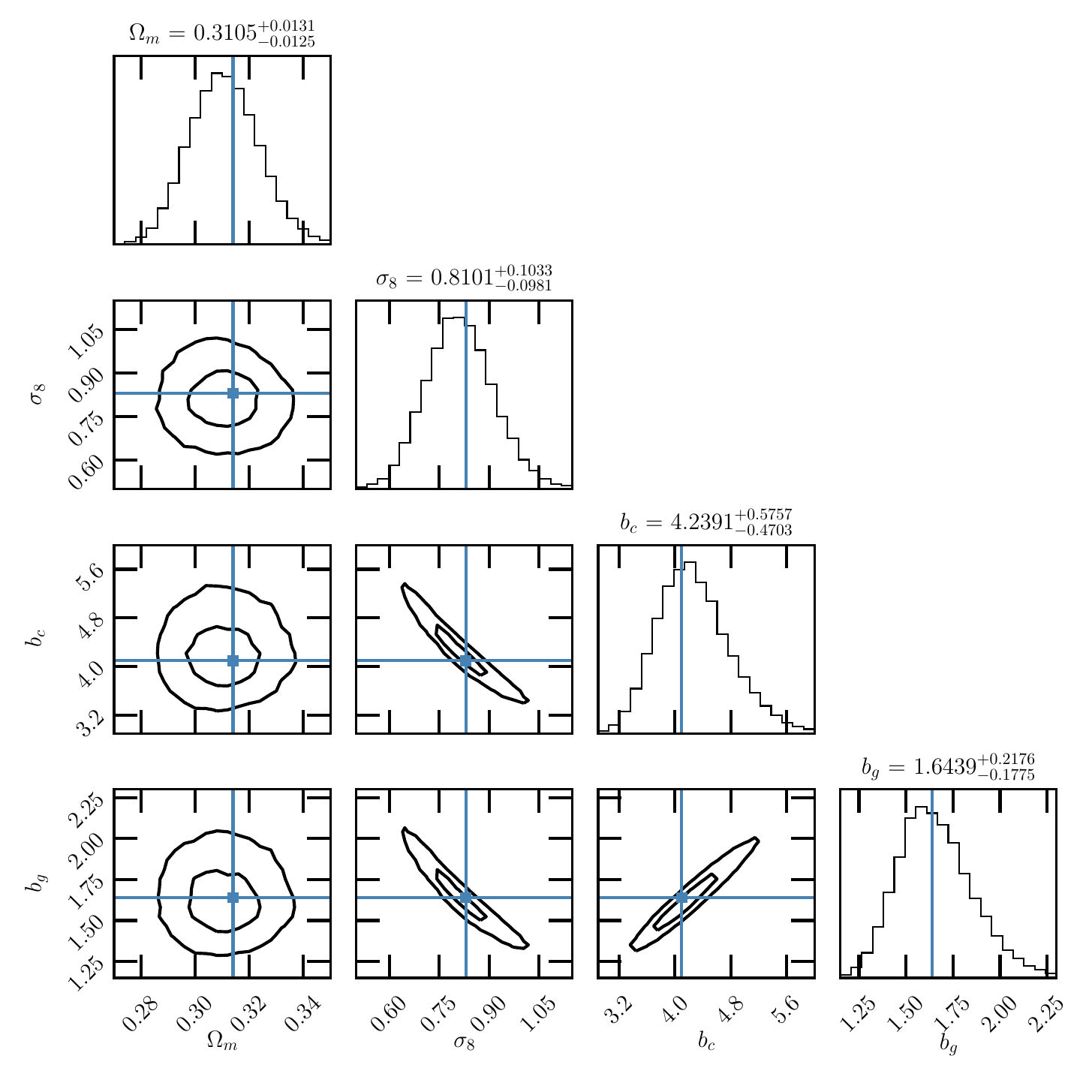}
\caption[]{Posterior distribution of cosmological parameters ($\OmegaM$ and $\sigma_8$) and nuisance parameters (cluster bias $\bc$ and galaxy bias $\bg$), derived from fitting projected cross-correlation functions $\wpcm$, $\wpcg$, and $\wpgg$. 
The contours correspond to 68\% and 95\% levels. 
The blue lines refer to true parameter values.  All contours capture the true values in the 68\% level. The calculated $\chi^2$ per degree of freedom is small ($\approx$ 0.2) because the total volume of our simulations is approximately 70 times the survey volume, and the data vector is much less noisy than real data.}
\label{fig:mcmc}
\end{figure*}

The observational data vector consists of the projected correlation functions $\wpcm$, $\wpcg$ and $\wpgg$, averaged over 20 {\sc Abacus Cosmos} realisations (the right-hand panel in Fig.~\ref{fig:wp}): 
\beq
\vec{x}_\mathrm{obs} = \bigg\{\wpcm(\rp),\ \wpcg(\rp),\ \wpgg(\rp)\bigg\} \ .
\label{eqn:obs_vec}
\eeq
We use ten logarithmic-spaced $\rp$ bins between 10 and 100 $\hiMpc$.

The corresponding model data vector is calculated with 
\beq
    \vec{x}_{\mathrm{model}} = 
    \bigg\{\bc \wpmm(\rp),\ \bg^2 \wpmm(\rp),\ \bc \bg \wpmm(\rp)\bigg\} \ ,
    \label{eqn:model_vec}
\eeq
where $\wpmm$ is calculated using the linear matter power spectrum $P(k)$ as calculated by {\sc{CAMB}} \citep{Lewis00} for a given set of cosmological parameters.

To calculate the covariance matrix of $\wpcm$, we first calculate the covariance matrix of $\Delta\Sigma$ based on the approach presented in \cite{Wu2019}. The covariance is dominated by shape noise at small scales and large-scale structure noise at large scales.  We assume a DES-like survey condition: a sky coverage of 5000 square degrees, clusters at $0.2 < z_\mathrm{lens} < 0.35$ (corresponding to a comoving volume 0.37 $h^{-3}{\rm Gpc}^3$) and above $2 \times 10^{14}~\hiMsun$, source galaxies at $z_\mathrm{source} = 0.75$ with a surface density 10 arcmin$^{-2}$.  To convert from the covariance matrix of $\Delta\Sigma$ to that of $\wphm$, we apply the linear transformation presented in \cite{Park21local}.   We note that the $\bc$ affects the lensing noise, and we use the bias corresponding to haloes of $2 \times 10^{14}~\hiMsun$ instead of the abundance-matched clusters because the former is closer to our $\bc$ from mock catalogues.

To compute covariance matrices for $\wpcg$, $\wpgg$ and their cross-term, we use the Gaussian analytic formalism found in \citet{Salcedo20}; also see e.g.~\citet{Marian15} and \citet{Krause17}. We again assume a DES-like survey condition and use non-linear power spectra calculated from our simulations.  We show the fractional error of $\wpcm$, $\wpcg$ and $\wpgg$ in Fig.~\ref{fig:frac_err} and the correlation matrix in Fig.~\ref{fig:corr}.  Selected diagonal values are shown in the correlation matrix.

With the ingredients above we can calculate the $\chi^2$ for different model vectors by 
\beq
 \chi^2 = 
(\vec{x}_\mathrm{obs} - \vec{x}_\mathrm{model})^T \mathcal{C}^{-1}
(\vec{x}_\mathrm{obs} - \vec{x}_\mathrm{model}),
\label{eqn:chisq}
\eeq
where $\mathcal{C}^{-1}$ is the inverse of the combined covariance matrix of $\wpcm$, $\wpcg$ and $\wpgg$.  This $\chi^2$ is used as the negative two times the log-likelihood function in the MCMC calculation.

\subsection{Parameter inference}

We perform a likelihood analysis to constrain $\OmegaM$, $\sigma_8$, $\bc$, and $\bg$ using the parallel affine-invariant ensemble sampler \citep{GoodmanWeare10} implemented in the Python module {\sc emcee}\footnote{We use 
{\sc emcee} 3.1.1, 
{\sc Corrfunc} 2.4.0, 
{\sc camb} 1.3.2, and 
{\sc corner} 2.2.1 in our calculations.} \citep{emcee}.

We initialise 200 walkers uniformly using the initialisation range listed in Table~\ref{tab:prior}.
The initialisation of walkers enables the parallelisation of the code, where processors handle multiple walkers simultaneously. The small range of initialisation does not limit the exploration range of the sampler because walkers quickly branch out and reach the rest of the parameter space. We assume flat priors that are listed in the `Prior Range' column of the table. The final chain has 184k steps in total, and we remove the first 18k as the burn-in. The chain was stopped according to the integrated autocorrelation time criteria. At the end of the chain, the ratio between the number of samples and the autocorrelation time is 22.

Fig.~\ref{fig:mcmc} shows our posterior distribution of parameters, generated using the {\sc Corner} software package \citep{corner}.  The contours refer to 68\% and 95\% boundaries, and the blue vertical and horizontal lines refer to fiducial parameter values.  All contours capture the true values in the 68\% level.  We compare various $\chi^2$ values:
\begin{itemize}
    \item $\chi^2_{\mbox{fid}}$ = 5.583,
    \item $\chi^2_{\mbox{best-fit}}$ = 3.380,
    \item $\chi^2_{\mbox{median}}$ = 3.421.
\end{itemize}
The degrees of freedom are 26, and thus the $\chi^2$ per degree of freedom is much less than 1.  Since the total volume of our simulations is approximately 70 times the survey volume, we expect the data vector to be much less noisy than real data, and thus the $\chi^2$ per degree of freedom is small.

Overall, we see that $\OmegaM$ is constrained at the 4.1\% level and that $\sigma_8$ is constrained at the 12.4\% level.  The modest constraint on $\sigma_8$ is due to the strong degeneracy with $\bc$ and $\bg$, both constrained at the 10\% level. In comparison, in Table 5 in \cite{Salcedo20}, the row corresponding to the large-scale (3 $\hiMpc$) $\DS$, $\wpcg$, and $\wpgg$ leads to a 3.7\% constraint on $\sigma_8$ with fixed $\OmegaM$.  Our constraints on $\sigma_8$ are weaker due to the larger scale cut and the free $\OmegaM$.  This comparison highlights the benefit of a stronger prior on $\OmegaM$ and smaller scale cuts.

We note that the contour of $\sigma_8$ and $\OmegaM$ does not show the usual anti-correlation from cluster number counts \citepalias[e.g.][]{DESY1CL}.   
Fig.~8 in \cite{Salcedo20} shows that the $\sigma_8$ and $\OmegaM$ have the opposite effect in determining $\wpgg$ and $\wpcg$; that is, their derivatives with respect to $\sigma_8$ and $\OmegaM$ have opposite signs.  Therefore, the constraints from  the cluster and galaxy clustering signals are highly complementary to the constraints from cluster abundance and lensing.

\begin{figure}
\centering
\includegraphics[width=1.0\columnwidth]{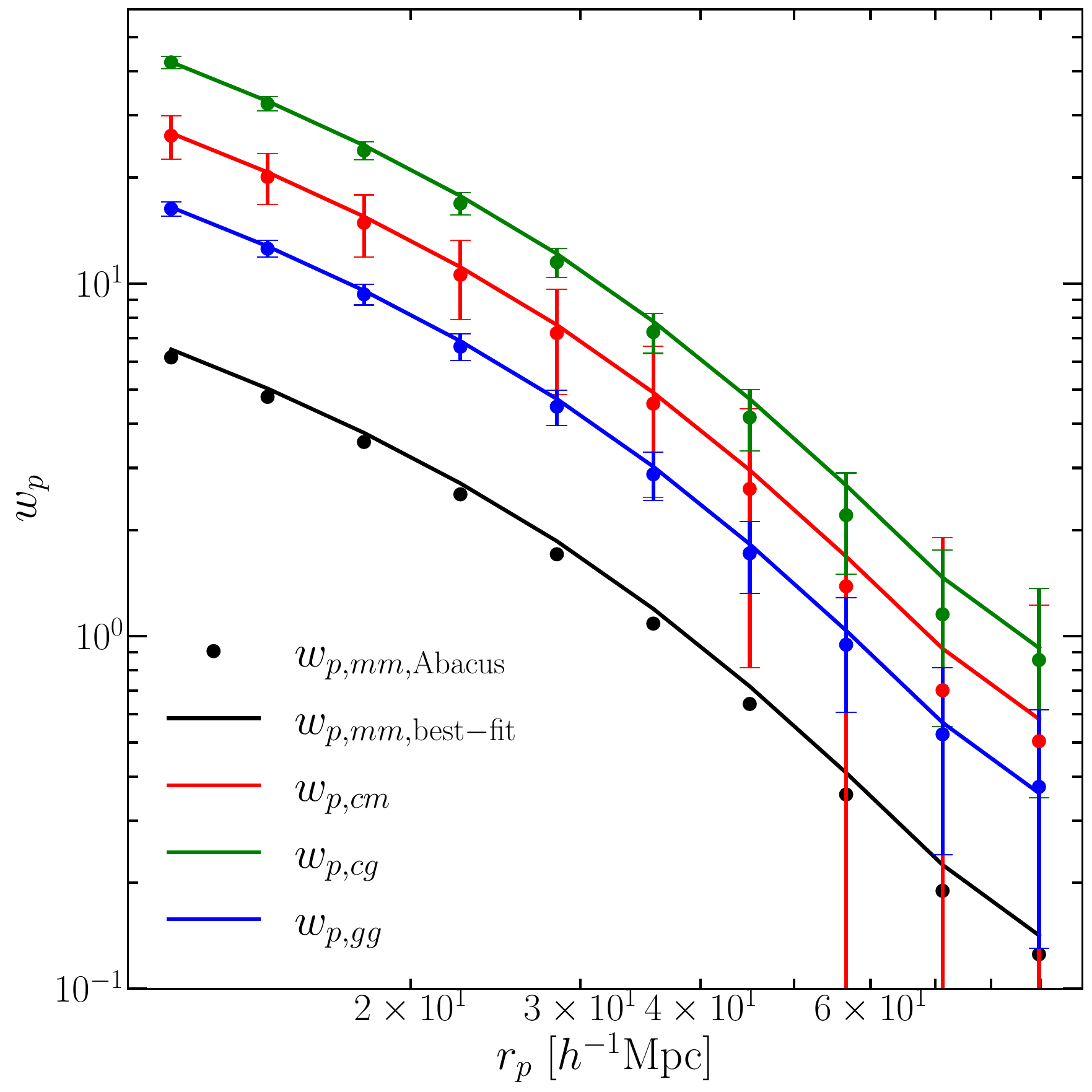}
\caption[]{The best-fit model (colour curves) compared with the input data vector from our mock catalogues (colour points with error bars). The black curve shows the analytic $\wpmm$ from the best-fit parameters, and the black points show the  $\wpmm$ from mocks.}
\label{fig:best-fit}
\end{figure}

In Fig.~\ref{fig:best-fit} we present the best-fit model, together with the data vector and its uncertainties. Since the uncertainties increase with scale (also see Fig.~\ref{fig:frac_err}), the best-fit model is mostly driven by the smallest $\rp$ bins.  At larger scales, the $\wpmm$ predicted by the best-fit parameters shows a small excess compared with the $\wpmm$ from mock.  This is related to the small excess shown in Fig.~\ref{fig:wp_compare} and the slightly larger $\OmegaM$ and $\sigma_8$ compared with the true values.  This excess is much smaller than the current level of experimental uncertainties but would require further examination for future data sets.

\begin{table}
	\centering
	\caption{Parameters in our likelihood analysis. We show the fiducial values, initialisation ranges, assumed priors, and best-fit values.}
	\label{tab:prior}
	\begin{tabular}{ccccc}
		\hline
		Parameter & Fiducial Value & Initialisation & Prior Range & Best-fit\\
		\hline
		$\OmegaM$ & $0.314$ & $[0.30, 0.32]$ & $[0.20, 0.40]$ & $0.310$\\
		$\sigma_8$ & $0.830$ &  $[0.80, 0.85]$ & $[0.50, 1.00]$ & $0.840$\\
		$\bc$ & $4.097$  & $[3.80, 4.20]$ & $[3.00, 6.00]$ & $4.106$\\
	    $\bg$ & $1.639$ & $[1.50, 1.80]$ & $[1.00, 3.00]$ & $1.591$\\
		\hline
	\end{tabular}
\end{table}

\section{Discussions}
\label{sec:discussion}

In this section, we discuss our results in the context of previous studies.  We will then describe our plans for further developing the model and applying our method to real data.

\subsection{Comparison with previous studies}

Using N-body simulations, \citet{Osato18} have shown that cluster surface density profiles exhibit a strong dependence on the orientation with respect to the line of sight \citep[also see e.g.][]{Dietrich14, ZZhang22}.  They have shown that this orientation dependence extends to 100 $\hiMpc$ and can be explained by the anisotropic halo--matter correlation function $\xihm(s,\mu)$.  For a mass-selected halo sample, we expect that averaging over all haloes and all $\mu$ would recover the isotropic $\xihm(r)$.  The fact that we find $\xicm$ higher than $\xihm$ indicates that the richness selection prefers clusters with  more strongly anisotropic $\xihm$ (e.g. due to the filaments along the line of sight).  Our finding is consistent  with their results of non-vanishing large-scale selection bias due to projection. 

Using mock cluster catalogues constructed from a HOD model, \cite{Sunayama20} have demonstrated that the cluster lensing and cluster clustering signal are boosted
relative to an isotropic halo model. Such a boost persists to large scales.  Their Fig.~13 shows that clusters that suffer from strong projection effects exhibit a highly anisotropic projected correlation function, indicating the existence of line-of-sight filaments.   

In our previous work \citep{Wu22}, we have studied the cluster projection effects using the mock \redmapper catalogues constructed from the Buzzard simulations, which are designed for DES mock analysis.  We have used the full dark matter particles from the simulations and have focused on relatively small scales ($<$ 3 $\hiMpc$ for $\xihm$ and $<$ 30 $\hiMpc$ for $\wpcm$).  Those results have hinted at a vanishing selection bias for both $\xihm$ and $\wpcm$ at large scales.  In addition, the Buzzard simulations have a lower cluster abundance compared with observed clusters.

In this paper, we focus on a regime complementary to \citet{Wu22}. We construct mock catalogues with a simple yet realistic HOD that matches DES cluster abundance, use large-volume N-body simulations, and focus on large-scale correlation functions.  We have found that the large-scale selection bias is non-vanishing and approaches a constant for $\xicm$ and $\wpcm$. In particular, the $\xicm$ selection bias is associated with projection effects, which is confirmed by calculating $\xicm$ in various orientations. 

The measurements of cluster clustering have recently come to fruition due to the availability of large-area survey data \citep[e.g.][]{Chiu20, To21b, Park21}. These analyses have considered or incorporated cluster selection bias in various ways.  
\cite{Chiu20} use the auto- and cross-correlation functions ($\xicc$, $\xicg$, and $\xigg$) between Hyper Suprime-Cam's CAMIRA cluster catalogue and the CMASS galaxy catalogue to constrain the normalisation of the richness--mass relation.  They have assessed the impact of selection bias and concluded that it is unimportant for their data set but would be necessary for future studies.
\cite{To21b} combine \redmapper cluster abundance with the auto- and cross-correlation functions between clusters, galaxies, and weak lensing shear.  They focus on the angular correlation function $w(\theta)$ and find that the selection bias is at the 15\% level at scales greater than 8 $\hiMpc$.
\cite{Park21} use cluster abundance, cluster lensing, and cluster clustering of the SDSS \redmapper catalogue and apply an empirical model for the projection effect.  They have found a 15--20\% anisotropic boost, similar to that in \cite{To21b}. They have found a lower $\OmegaM$ and higher $\sigma_8$ compared with the {\em Planck} results.

Our results imply that the projection effects impact not only the projected correlation functions but also the 3D correlation function.  In both cases, we can self-consistently model the large-scale cross-correlation functions between clusters, galaxies, and shear and use them to solve for the selection bias.  We have not considered constraints from small-scale correlation functions, which are more difficult to model but have enormous constraining power \citep[][]{Salcedo20, Salcedo22}.  Modelling the small-scale correlation function would require extra nuisance parameters, which may weaken the constraining power.  We expect that the large-scale self-calibration we present in this work would be highly complementary to the small-scale bias calibration.

\subsection{Future work}

In this work, we focus on cluster selection bias and simplify the assumptions on other systematic uncertainties.  In particular, we use the 3D positions of galaxies in simulations and ignore galaxy velocities and redshift uncertainties.  The redshift uncertainties of clusters are likely to remain negligible, but the photometric redshift uncertainties of galaxies need to be modelled.  A full analysis would need to take into account the photometric redshift uncertainties \citep[e.g.][]{ZWang_et_al_2019} and the redshift-space distortion \citep[e.g.][]{Kaiser87,  Hamilton1998, Sunayama22}.

We assume that the galaxy sample and the cluster sample are constructed from galaxies with different colour selection criteria, and therefore we need two separate HOD models.  The different colour selection criteria are due to the fact that the galaxy sample and the cluster sample are optimised differently: the galaxy sample is optimised for a small redshift uncertainty, while the cluster sample is optimised for a small richness--mass scatter \citep{Rykoff14, Rozo16}.  However, the different colour selection criteria lead to a large number of nuisance parameters, which could be difficult to constrain.  We plan to explore the possibility of using the same colour selection criterion for both samples, which will require only one set of HOD parameters. This approach would potentially optimise both samples simultaneously and improve their constraining power on HOD parameters. 

With the upcoming spectroscopic galaxy sample from the Dark Energy Spectroscopic Instrument (DESI) and Nancy G.~Roman Space telescope, it is possible to cross-correlate cluster samples with spectroscopic galaxies \citep[analogous to][]{Gazta2012}. We expect that the spectroscopic galaxy sample would have a smaller sample size but a better-constrained HOD. 

With the newly available large-area multi-wavelength cluster samples, it is also possible to cross-correlate optical galaxies with clusters selected by X-ray or the SZ effect.  For example, \cite{Shin21} measure cluster lensing and galaxy clustering around clusters selected by SZ-signal from the Atacama Cosmology Telescope.  Compared with optical cluster samples, X-ray and SZ cluster samples focus on more massive haloes and have smaller sample sizes.  Another approach would be using clusters with various mass proxies and performing both self- and cross-calibration of cluster selection bias \citep[see e.g.][]{Costanzi21}.

In this work, we use the large-scale correlation function to demonstrate the feasibility of self-calibrating selection bias.  On the other hand, small-scale correlation functions have enormous constraining power \citep[see][for detailed discussion]{Salcedo20, Salcedo22}.  The modelling of small-scale correlation functions would require detailed simulations covering a wide range of parameters.  These simulations are usually computationally extensive, but the recent development of emulators provides an effective approach for constructing small-scale models \citep[see e.g.][]{Nishimichi19, Wibking20}.  We plan to apply an emulator approach to accurately model the small-scale correlation functions and their dependence on galaxy--halo connection models.   

Galaxy clustering measurements suffer from various systematic uncertainties \citep[see e.g.][and references therein]{WeaverdyckHuterer21}.  For example, \cite{Pandey22} analyse DES Y3 \redmagic galaxy clustering and galaxy-galaxy lensing and find that the galaxy bias derived from galaxy clustering is systematically higher than the galaxy bias derived from galaxy--galaxy lensing. They parameterise this discrepancy by a decorrelation parameter $X_{\rm lens}$ and find that such a decorrelation can be alleviated by broadening the colour selection of the galaxy sample.  This result indicates colour-dependent systematic uncertainties in the galaxy catalogue, which need to be taken into account in cluster--galaxy cross-correlation studies.

Future photometric surveys like LSST demands more stringent control of systematic uncertainties compared with DES-like surveys \citep[see e.g.][for a review]{Mandelbaum18}.  For example, the blending of galaxies would become more significant, impacting the galaxy shape and photometric redshift measurements.  These systematic uncertainties are likely to be resolved by the cross-calibration between LSST, Roman, and Euclid, as well as spectroscopic follow-up observations from the ground \citep[e.g.][]{Rhodes17, Eifler20WFIRSTLSST}.

\section{Summary}
\label{sec:summary}

Using mock catalogues of galaxies and galaxy clusters based on N-body simulations and HOD models, we assess the efficacy of using cluster lensing, cluster--galaxy cross-correlation functions, and galaxy auto-correlation functions to self-calibrate the optical cluster selection bias.  Although cluster selection bias is mostly due to projection effects, we have found that the selection bias is present even in 3D correlation functions and extends to $\approx 100 ~ \hiMpc$.  Using the 2D correlation functions, $\wpcm$, $\wpcg$, and $\wpgg$, we show that the selection bias can be calibrated self-consistently at scales larger than $10 ~ \hiMpc$ (Fig.~\ref{fig:wp}).  We perform a likelihood analysis using a data vector derived from simulations and analytic covariance matrices assuming a DES-like survey condition (5000 deg$^2$, 10 source galaxies per arcmin$^2$, and focusing on the lowest redshift bin $0.2 < z < 0.35$ and large scale 10 -- 100 $\hiMpc$).  We find that $\OmegaM$ and $\sigma_8$ are constrained at the 4.1\% and 12.4\% levels respectively and exhibit only mild degeneracy. The cluster bias $\bc$ and galaxy bias $\bg$ are strongly degenerate with each other and are constrained at 10\%.

The constraints forecasted here are modest due to the conservative scale cuts we use. We have discussed strategies for pushing the modelling to small scales and applying the method to real data.  Optical cluster cosmology is at a crossroads because of the newly uncovered systematic biases.  The success of future cluster experiments would likely require a concerted effort of self- and cross-calibrations of cluster selection bias.

\section*{Acknowledgements}

We thank Lehman Garrison and the {\sc Abacus} team for providing {\sc Abacus Cosmos} simulation suite. We thank Tomomi Sunayama, Chun-Hao To, and the anonymous reviewer for their helpful suggestions.  
During the preparation of this work, CZ and CMH are supported by David \& Lucile Packard Foundation award 2021-72096, the Simons Foundation award 60052667, and the NASA award 15-WFIRST15-0008.
HW is supported by the DOE award DE-SC0021916 and the NASA award 15-WFIRST15-0008. 
ANS is supported by the DOE awards DE-SC0009913 and DE-SC0020247.

The computations in this paper were performed on the CCAPP condo of the Pitzer Cluster at the \cite{OSC}.  We thank the developers for the following software packages: 
{\sc CAMB}, 
{\sc emcee} \citep{emcee},
{\sc Corner} \citep{corner}, 
and {\sc Corrfunc} \citep{Corrfunc}.

\section*{Data Availability}

The data underlying this analysis will be shared upon reasonable request to the corresponding author.

\bibliographystyle{mnras}
\bibliography{references_cz} 

\bsp	
\label{lastpage}
\end{document}